\documentclass{osa-article}

\journal{osac}

\articletype{Research Article}

\begin{document}

\title{Excitation of Symmetry Protected Modes in a Lithium Niobate Membrane Photonic Crystal for Sensing Applications}

\author{Ayman Hoblos,\authormark{1,*} Miguel Suarez,\authormark{1} Nadège Courjal,\authormark{1} Maria-Pilar Bernal,\authormark{1} and Fadi I. Baida\authormark{1}}

\address{\authormark{1}Institut FEMTO-ST, UMR CNRS 6174, Université Bourgogne Franche-Comté, 25030 Besançon cedex, France\\}

\email{\authormark{*}ayman.hoblos@femto-st.fr} 



\begin{abstract}
Our theoretical study reveals the opportunity to develop an electric field sensor based on the exploitation of Symmetry Protected Mode (SPM) that we excite within an electro-optical material, namely lithium niobate (LN). The SPM consists of a dark Fano-like resonance that results from the combination of a discrete Bloch mode of a Photonic Crystal (PhC) with a continuum mode of a membrane, both of them made in LN. The dark character is linked to the structure geometry having a high degree of symmetry. The SPM excitation is then made possible thanks to an illumination under small oblique incidence which breaks the symmetry of the configuration. This results in several ultra-sensitive and tunable Fano-like resonances with high quality factors up to $10^5$ in the telecoms spectral range. Some of these resonances provide modes with a highly confined electric field inside LN. This confinement allows the enhancement of the electro-optic Pockels effect by a factor up to $4 \times 10^5$, thus exacerbating the detection sensitivity of the designed sensor.
\end{abstract}

\section{Introduction}
Today, the world faces a large number of challenges in different disciplines such as health, data processing, energy, information flow and environment. The growing demand for high sensitivity of detection in medical, military and other applications, has led to the emergence of new and impressive solutions to detect weak signals. Photonic technologies and in particular Photonic Crystals (PhCs) are one of the most important solutions in sensing applications. They are now widely used for the design of electric field sensors (E-field sensors) \cite{calero:sr19}, magnetic field sensors \cite{cetin:bb19,saker:jce19}, temperature sensors
\cite{wang:ao19, arumuru:sr19,qiu:ol16}, gas-sensors \cite{sunner:apl08,liu:sr19} and other biological sensors \cite{obeid:bb17,zhou:oex19}. Several physical phenomena can be exploited for the detection of these quantities. Let us quote for example the piezoelectric effect \cite{zhang:jpcc15}, the photorefractive effect \cite{kostritskii:oqe20}, the conductive and inductive effects \cite{putley:ao65,logothetis:apl72}, the pyroelectric effect \cite{wang:am16}... In optics, and to avoid the use of very powerful light sources, it is often convenient to rely on a linear effect. The Pockels effect perfectly fulfills this condition because it allows, through an electro-optical (EO) material, to induce a change in the optical properties of the material that is linearly linked to the E-field to be measured. We focus in this paper on the E-field sensors based on PhC engraved within highly efficient EO medium.

However, when compared with traditional electronic technology, E-field sensors based on metal-free structures offer several advantages because of their excellent electrical insulation \cite{wang:apl19,zeng:apl11}. They are much more compact, which results in a higher stability and less distortion to the E-field distribution \cite{ogawa:jlt99}.

 The conventional principle of an EO sensor is as follows: an external E-field induces the modification of the refraction index of the material. This modification leads to a redistribution of the electromagnetic field inside the material and, consequently, modifies its optical properties that are detected by reflection or by transmission. The key-point is then to design a configuration where this distribution is important enough to allow the detection of very weak signals. This can be achieved by highly confining the electromagnetic density of states \cite{Lourtioz:tonanophotondev05}. 

In this context, several studies were conducted showing different configurations with a high aptitude to detect weak signals. Roussey \textit{et al.} \cite{roussey:apl05} present a simple structure of PhC made of air holes engraved into a lithium niabate (LN) waveguide. They show a theoretical and experimental photonic band gap shift $300$ times larger than expected through a non-structured waveguide (without PhC). To bypass the experimental drawbacks due to the light injection into waveguides, Qiu \textit{et al.} \cite{qiu:opex16} proposed to excite membrane PhC Fano resonances at normal incidence. In that study, Fano resonance is obtained by coupling a discrete Bloch mode of a PhC with a continuum mode of a LN membrane. They got a sensitivity of $\Delta\lambda_{res}/\Delta E=0.25\times{10}^{-5}$ nm.m/V.

As in the studies cited above, we will choose the LN as an active material, due to its high EO coefficient with high transparency in the visible and infrared wavelength regime \cite{lipson:opex18,weis:apa85,guarino:nphoton07}. LN is well known for its high nonlinear optical coefficients, piezoelectric, ferroelectric and large acousto-optic coefficient \cite{wong:book02}.

Seeking a configuration exhibiting a resonance with high quality factor, we came across the symmetry protected mode (SPM) \cite{elganainy:np18,lu:np16,yoon:sr15}. This kind of resonance provides an excellent electromagnetic field confinement essential to enhance the LN EO effect surpassing a conventional Fano resonance \cite{qiu:opex16}. As it is well-known \cite{zhang:prb10,campione:acs16}, SPM generally occurs when a structure with a high degree of symmetry is assumed. The excitation of such modes needs a small break of this symmetry which has an important impact on their resonance Q-factor. For example, a square grating photonic crystal with an elementary pattern, having a perfectly symmetrical shape with an axis of symmetry parallel to the incident beam, is an excellent candidate for exhibiting such modes. At normal incidence, these modes still are dark ones. At oblique incidence, such SPM turns into bright modes and can be easily excited. The dark character of such modes can be seen as a  destructive interference (quality factor tends to infinity) between more than two degenerated modes. SPMs have recently been used to demonstrate high-quality factor resonances near normal incidence (more than $10^9$) that may be exploited for various applications \cite{lee:prl12,kodigala:nat17}. Therefore, motivated by the SPM, we propose to excite such dark modes by introducing a small symmetry breaking through the illumination direction.

Our theoretical studies reveal the opportunity to develop an E-field sensor based on a tunable resonance with a high Q-factor (up to $1.2 \times 10^5$) and a robust extinction factor of $60\%$. We reach a new horizon with our structure due to an intensity enhancement that locally reaches $60000$. This allows a detection sensitivity of $4$ pm.m/V, $160000$ times greater than the sensitivity obtained through a Mach-Zehnder interferometer design \cite{lipson:opex18} and $1600$ times greater than what was predicted with a membrane PhC Fano resonance \cite{qiu:opex16}.

\section{The Proposed Structure}

First, as in refs. \cite{qiu:opex16,calero:sr19}, we choose a configuration based on a PhC operating near the $\Gamma$ direction (out-off-plane illumination around normal incidence). For an efficient detection, the PhC period must be smaller than the operation wavelength providing only one direction (only the diffracted zero-order is propagative) for the reflected and transmitted energy. This condition is fulfilled, at normal incidence illumination when  $p < \lambda$, with p being the period of the structure and $\lambda$ is the operating wavelength. 
\begin{figure}[h!]
	\includegraphics[width=120mm]{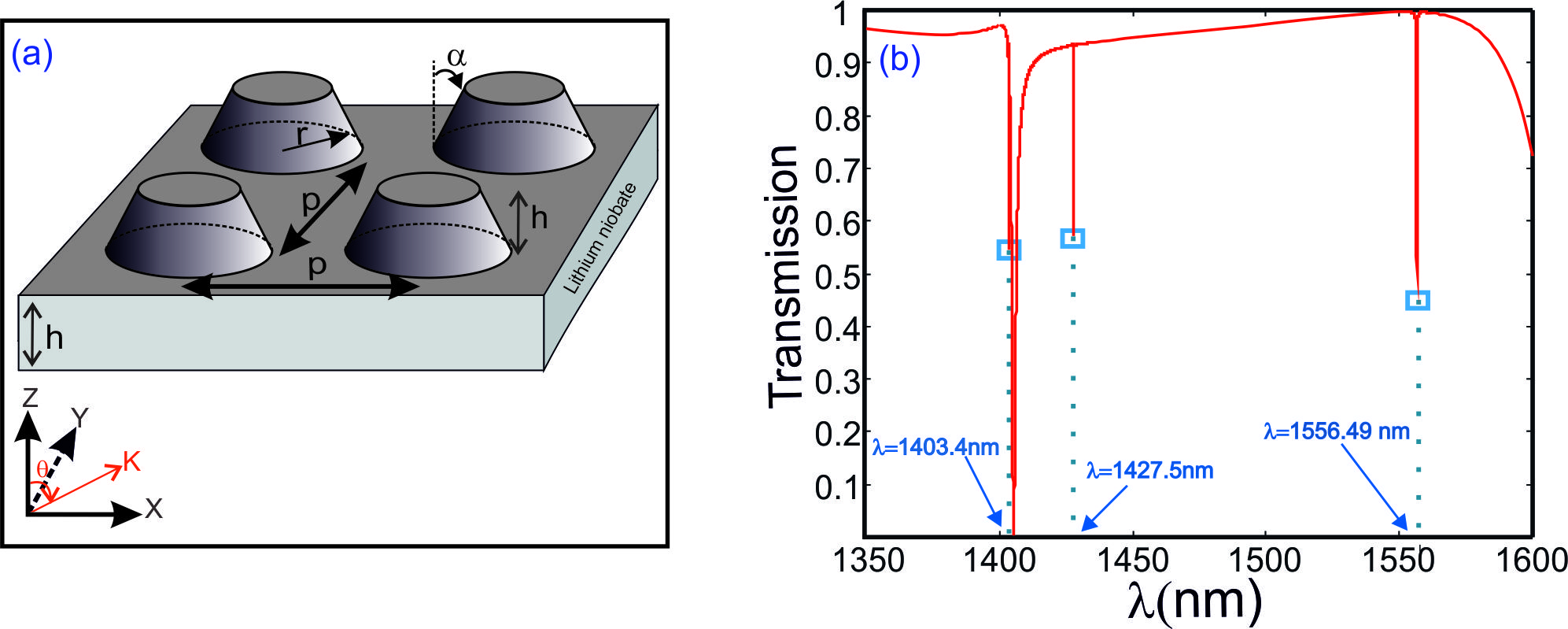}
	\centering
	\caption{(a) Schematic of the proposed structure, (b) the transmission spectrum of the complete structure.}
	\label{fig:struc2}
\end{figure} 

To excite Fano SPM resonances, we propose a PhC, which supports spectrally discrete Bloch modes, to be combined with a membrane which allows the existence of broadband guided modes. If we illuminate the membrane at normal incidence, no guided modes can be excited due to the perfect axial symmetry (counter propagating modes). Nonetheless, as shown on Fig. \ref{fig:struc2}(a), the cylindrical symmetry of the PhC patterns together with the square symmetry (C4 axis) of the Bravais lattice keeps prohibited the excitation of the guided modes at normal incidence. Fortunately, an oblique incidence can lift this ban and lead to the excitation of some dark guided modes. The signature of this excitation is then a sharp resonance (dip for instance) in the transmission spectrum of the structure (see Fig. \ref{fig:struc2}(b)). Moreover, working with a small angle of incidence yields to a small break in the symmetry of our structure.  The smaller the angle of incidence, the greater the Q-factor of the resonance.

 Consequently, our structure relies on a square lattice PhC of period $p_x=p_y=p=1050$ nm. The unit cell consists of a truncated conical rod with a base's radius $r = 0.3\times p$, a height $h=p/3=350$ nm and a conicity angle $\alpha=15^o$ engraved within a LN layer of $2\times h=700$ nm thickness as presented in the schematic of the structure in Fig. \ref{fig:struc2}(a). The fabrication of the structure has been studied previously in detail by G. Ulliac \emph{et al.} \cite{ulliac:me11}. The conical rods (PhC) are usually fabricated by Focused Ion Beam (FIB) milling, which is highly time consuming. G. Ulliac \textit{et al.} proposed an easy to implement batch technique consisting of writing these patterns by e-beam lithography followed by Reactive Ionic Etching.

The higher EO coefficient of LN is the $r_{33}$ one that involves the E-field component parallel to its optical axis. To benefit from this coefficient when illuminating the material at normal incidence, we should use a X- or Y-cut wafers of LN (the optical axis is then parallel to the electric incident field). Furthermore, as a first approximation, one can neglect the other EO coefficients of LN and write the optical refractive index modification induced by the presence of an external electric field $E_{s}$ through \cite{roussey:josab07}:
\begin{equation}
\Delta n(x,y,z)=-\frac{1}{2} \times n_{e}^{3} \times r_{33} \times E_{s}
\label{eq:dn}
\end{equation}
with $n_{e}=2.21$ is the extraordinary optical index of LN at $1.55$ $\mu$m. The structure is illuminated by a linearly polarized plane wave at oblique incidence (see angle $\theta$ in Fig. \ref{fig:struc2}(a)). The optical response is calculated using a home-made code based on the finite-difference time-domaine (FDTD) method \cite{Taflove2005} with an oblique incidence ($\theta = 0.5^\circ$). The transmission spectrum shows several sharp resonances (at least 4 resonances between $\lambda=1350$ nm and $\lambda=1600$ nm) as shown in Fig. \ref{fig:struc2}(b). 

In order to point out the origin of these resonances, we have studied the optical response of the PhC and the LN membrane separately. The obtained spectra (not presented here) show a single sharp Bloch mode resonance with a Q-factor $=5480$ for the PhC while an almost flat transmission coefficient is obtained for the membrane (continuum). Fortunately, by coupling the two structures, additional sharp resonances appear whose correspond to some dark modes (SPMs) that are excited thanks to the symmetry-breaking induced by the oblique incidence illumination.
\newline
The properties of the resonances in the spectral range $\lambda\in$[$1350$ nm;$1600$ nm] are summarized in table $1$ in terms of Q-factor and Extinction Ratio. The latter is defined by:  
\begin{equation}
\textrm{Extinction Ratio}=\frac{T_{max}-T_{min}}{T_{max}} 
\label{eq:exratio}
\end{equation}

where $T_{max}$ is the maximum transmission value just near the resonance dip and $T_{min}$ is the minimum transmission value at resonance. Note that a high Q-factor generally corresponds to high electromagnetic density of state, which is consisting with the origin of SPMs.

\begin{table}[]
\label{tab:table1}
\begin{center}
\caption{The properties of the four resonances of the transmission spectrum presented in Fig. \ref{fig:struc2}(b).}
\begin{tabular}{|c|c|c|c|}
\hline
\textbf{Spectral position} (nm) & \textbf{Q-Factor} & \textbf{Extinction Ratio} (\%) \\
\hline
$\lambda_1^{res}=1403.56$  & 118339    & 57    \\
\hline
$\lambda_2^{res}=1405.35$   & 860    & 99    \\
\hline
$\lambda_3^{res}=1427.71 $  & 101631    & 50.61    \\
\hline
$\lambda_4^{res}=1556.59$    & 9612    & 55 \\ \hline
\end{tabular}
\end{center}
\end{table}

Otherwise, for $\lambda_2^{res}$, the relatively low Q-factor indicates a resonance with a slightly confined mode, which can't match an SPM. In the following, this mode will no longer be analysed since it is considered to be unsuitable for the EO detection application.

The transmission spectrum for the coupled structure is difficult to be quantitatively understood, but the origin of the SPM resonances is most likely linked to the interference of the PhC Bloch modes with the dark modes (guided modes) of the membrane or vice versa. This will be discussed in more details in the next section.

\section{Results and Discussion}

While some modes (bright modes) are excitable at normal incidence, others (dark modes) remain inaccessible due to structure symmetry. For out of normal incidence, the broken symmetry induced by this propagation direction triggers the selection rules and enables the excitation of SPMs. Moreover, the break of the symmetry can induce the degeneracy of some modes and remove it to some others. It makes dark modes become bright or vice versa (case of the Bound state In the Continuum (BIC) modes) \cite{yoon:sr15}.
\newline

\begin{figure}[h!]
	\centering
	\includegraphics[width=140 mm]{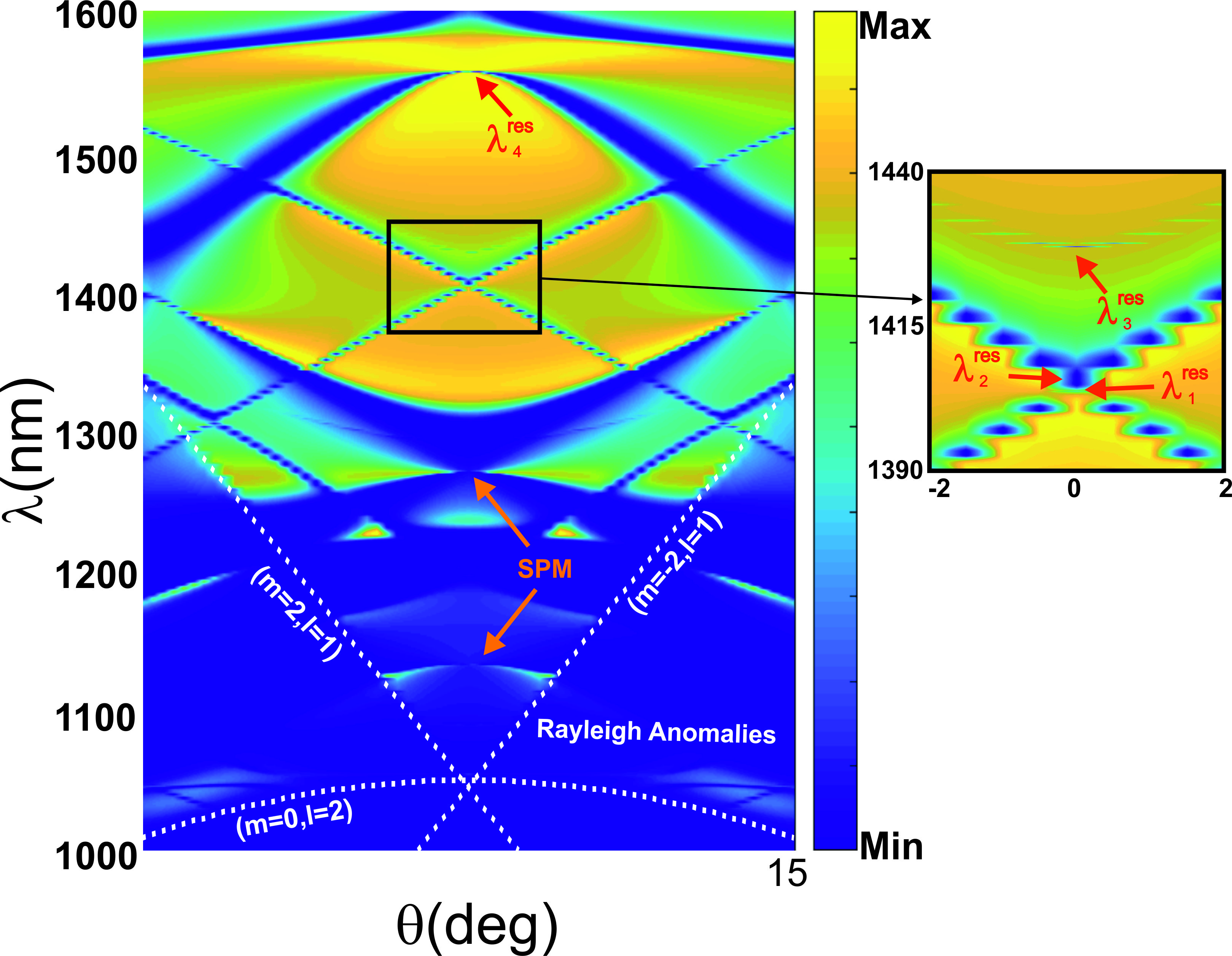}
	\centering
	\caption{Transmittance to the power of five ($T^5$ instead of $T$ to enhance the visibility of the different resonances) as a function of wavelength $\lambda$ and angle of incidence $\theta$ in TM-polarization. The white dashed lines correspond to the Rayleigh anomalies calculated using the Eq.1 in \cite{ndao:jo18}. The four resonances of table 1 are indicated in red arrows but other SPMs appear for smaller values of the wavelength (see ochre arrows). A zoom-in of a the part framed by a rectangle is made on the right part of the figure to point out the two close resonances $\lambda_1^{res}$ and $\lambda_2^{res}$. }
	\label{fig:TETM}
\end{figure}

To confirm the nature of our three SPMs, we present on Fig. \ref{fig:TETM} the angle-dependence transmission diagram calculated using the FDTD algorithm \cite{belkhir:physre08} for an incident TM-polarized plane wave. This diagram exhibits many sharp resonance features between $\lambda=1000$ nm and $\lambda=1600$ nm. One can clearly see the extinction of some resonances at some critical angles and the appearance of others. In a general way, the resonances marked by red arrows show vanishingly narrow transmission dips as $\theta$ approaches $0^\circ$, except $\lambda_{2}^{res}$. It shows an invarariant transmission behavior versus $\theta$. Therefore, with a small variation of $\theta$, one can achieve resonances with a huge Q-factor. By increasing the angle of incidence the resonances become too wide and the Q-factor gradually decreases \cite{yoon:sr15}. The resonance suppression at $\theta=0^\circ$ can be seen as destructive interference of emitted diffracted fields from two opposite sides of the symmetric structure. Let us mention that, due to the fabrication imperfections (residual roughness for instance), SPMs could be excited even at normal incidence. In that case, the Q-factor, that is related to the degree of the symmetry breaking, only depends on the preciseness (resolution) of the nano-fabrication process \cite{hoblos:ol20}.  \newline

\begin{figure}[h!]
	\centering
	\includegraphics[width=140 mm]{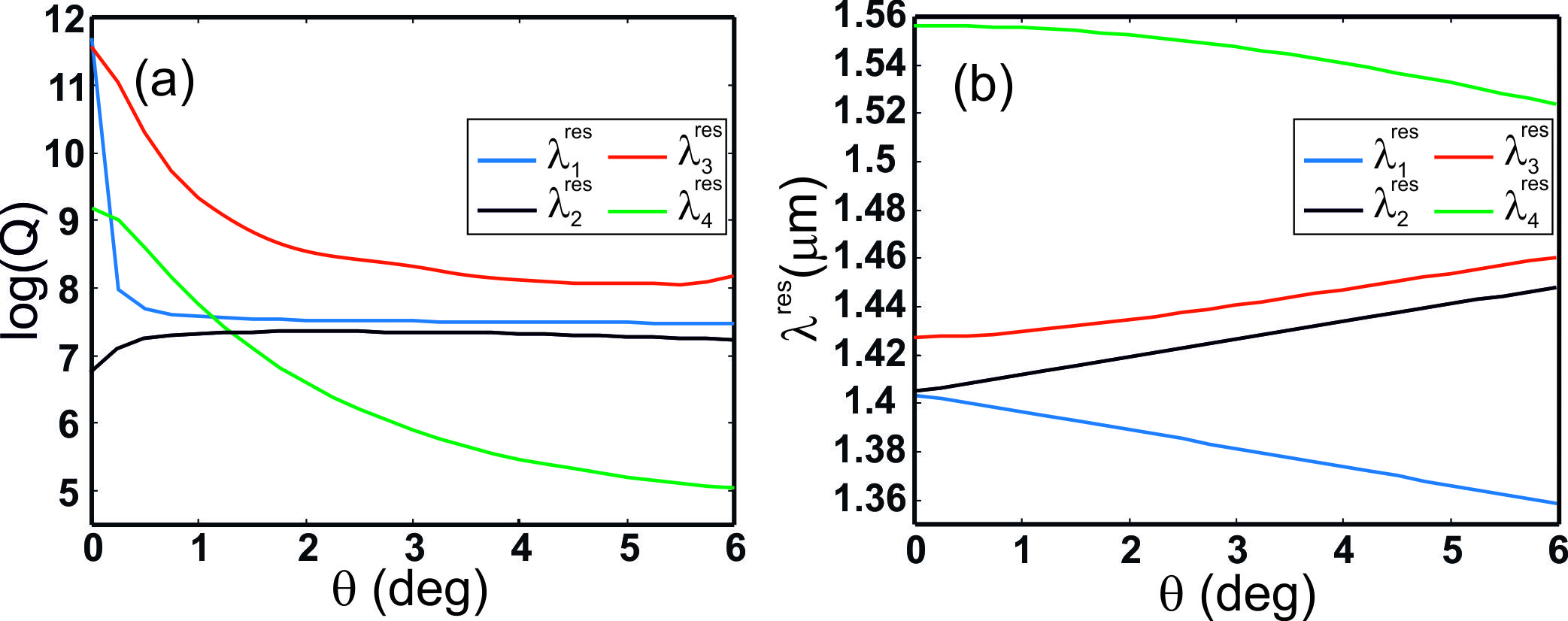}
	\centering
	\caption{(a): Evolution of the quality factor for different resonances in algorithmic scale with respect to the angle of incidence. (b): Evolution of spectral position of resonances with respect to the angle of incidence.}
	\label{fig:qfactor}
\end{figure}
In order to show more details about the SP properties of the four resonances indicating in Fig. \ref{fig:struc2}(b), we extract from the diagram of Fig. \ref{fig:TETM} their Q-factor and spectral position variations with respect to the angle of incidence as shown in Fig. \ref{fig:qfactor}. On the first hand, Fig. \ref{fig:qfactor}(a) denotes that the Q-factor of the resonances for $\lambda_1^{res}$, $\lambda_3^{res}$ and $\lambda_4^{res}$ exhibits a quite similar decreasing behavior when $\theta$ increases meaning a similar origin, namely SPM. Contrarily, for $\lambda_2^{res},$ a different behavior is obtained with a Q-factor that slightly grows but remains almost invariant for increasing angle of incidence (see the black triangles). On the second hand, Fig. \ref{fig:qfactor}(b) shows different behaviors for the spectral position with respect to the angle of incidence even for SPMs. More precisely, the spectral position for resonances at $\lambda_1^{res}$ and $\lambda_3^{res}$ increases with the angle of incidence $\theta$ while they do not have the same origin. So, the evolution of the spectral position of the resonances does not give a discriminating information on the origin of the resonance. \newline

In order to choose the most efficient resonance for our application (E-field sensing), we examine the influance of the electric field confinement on the electro-optical effect. The local expression of the latter is then derived from Eq. \ref{eq:dn} as \cite{qiu:opex16}:
\begin{equation}
\Delta n(x,y,z)=-\frac{1}{2} \times n_{e}^{3} \times f_{op}^{2}(x,y,z) \times r_{33} \times E_{s}
\label{eq:dnfop}
\end{equation} 
where $f_{op}(x,y,z)$ is the local optical field factor defined by
 \begin{equation}
f_{op}(x,y,z) = \frac{| \vec{E}(x,y,z) |_{structure}}{| \vec{E} |_{bulk}},
\label{eq:fop}
\end{equation}
where $| \vec{E}|_{bulk}$ is the amplitude of the homogeneous E-field in the bulk structure (without PhC) and $| \vec{E}(x,y,z)|_{structure}$ is the local amplitude of the E-field in the studied structure. Thus, it is essential to consider the SPM mode for which the $f_{op}$ is most important. Nonetheless, as the electro-optical effect is a property of the dielectric media, we are seeking resonance which corresponds to a "dielectric mode" for which the confinement of the E-field occurs inside the electro-optical material (LN for instance). When the confinement of the E-field is outside the structure (in the air), it is called "air mode".\newline

\begin{figure}[h!]
	\includegraphics[width=130mm]{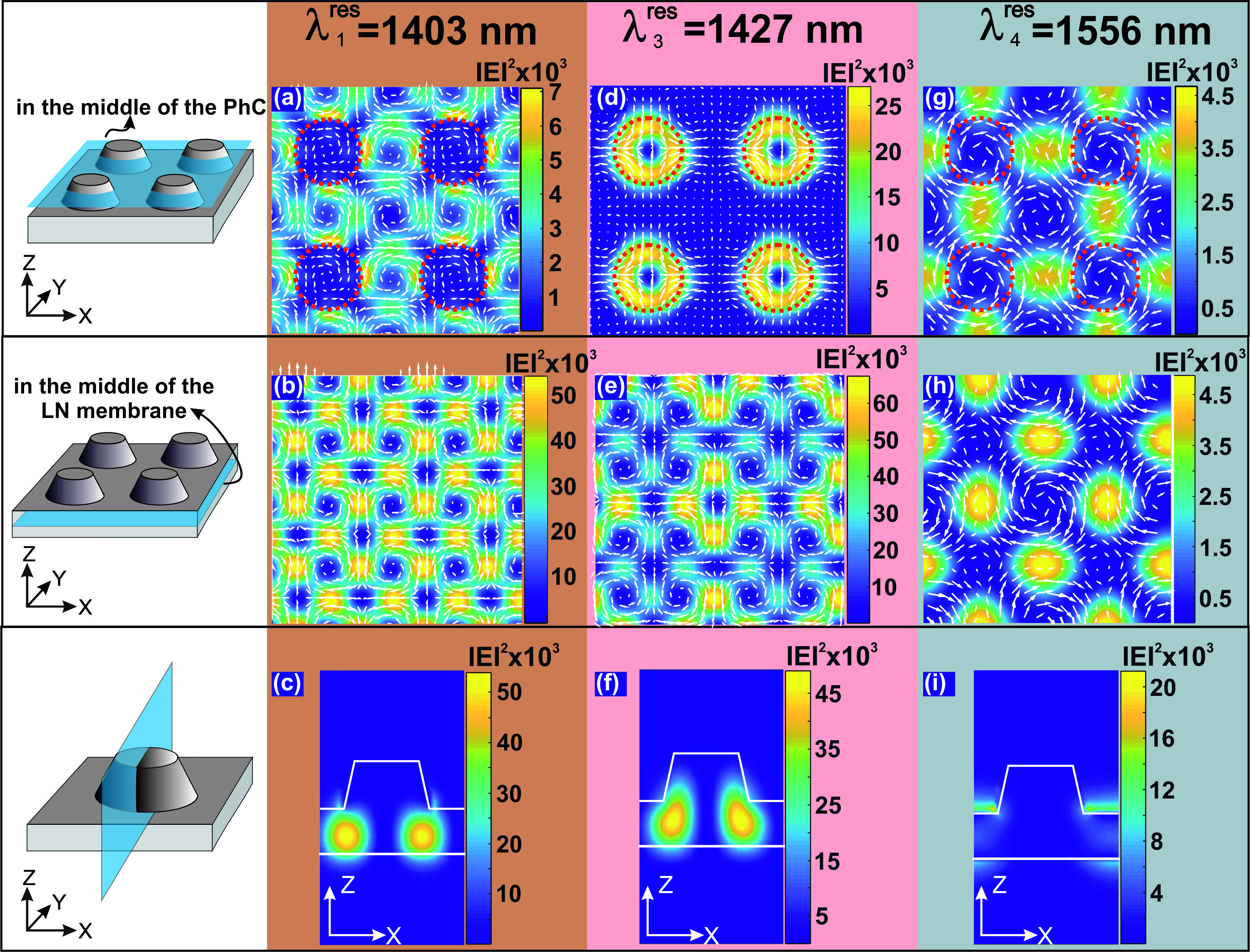}
	\centering
	\caption{Spatial distributions of $f_{op}^2$ for two $XY$-cuts and a $XZ$-cut at (a,b,c) $\lambda_1^{res}=1403$ nm, (d,e,f) $\lambda_3^{res}=1427$ nm and (g,h,i) $\lambda_4^{res} = 1556$ nm. Dashed lines show the structure edges and the white arrows indicate the tangential E-field vector. }
	\label{fig:distr}
\end{figure}

To fulfill these two conditions, we calculate the E-field distributions inside and around the structure for the three SPMs (at $\lambda_1^{res},\lambda_3^{res}$ and $\lambda_4^{res}$). Fig. \ref{fig:distr} show these distributions in $XY$ and $XZ$ plans associated to each resonant wavelength pointed by a blue square in Fig. \ref{fig:struc2}(b). 

The first column of Fig. \ref{fig:distr} shows the plans (in blue color) where the electric intensity is calculated. 
The second column of fig. \ref{fig:distr} corresponds to the resonance at $\lambda_1^{res} = 1403$ nm. From Fig. \ref{fig:distr}(a), that corresponds to the E-field intensity in a plan cutting the conical patterns at their mid-height, one can see the occurrence of three E-field vortices per period. The rotating character of the electric field (see the white arrows) denotes a symmetry which is inconsistent with a linearly polarized plane wave near normal incidence. This behavior appears even more clearly in Fig. \ref{fig:distr}(b) corresponding to a cross-section in the middle of the membrane. Here, eight electric intensity spots per period occur. These symmetries, different from the one of the linearly polarized incident wave, are the footprint of the SPM mode excitation. Fig. \ref{fig:distr}(c) demonstrates the dielectric-mode nature of the SPM at $\lambda=1403$ nm where the confinement clearly occurs inside the structure. In addition, the normalized E-field intensity reaches a maximum of $| E|^2 = 5 \times 10^4$, which corresponds to an optical local field factor $f_{op}(x,y,z) = 224$ making this mode highly desirable for EO applications. 
 
As far as $\lambda_3^{res} = 1427$ nm, we got a maximum of the normalized E-field intensity $|E|^2$ that exceeds $6 \times 10^4$ as shown in Figs. \ref{fig:distr}(d,e,f), which means an optical local field factor $f_{op}(x,y,z) > 245$. Furthermore, the electromagnetic mode is mainly located in the membrane with a small part in the  PhC, so it is also a dielectric mode. In the LN membrane (see Fig. \ref{fig:distr}(e), the E-field also shows vortices and high spots with corresponding to a higher degree symmetry incompatible with the axial symmetry of the incident wave nor with the C4 symmetry of the PhC. 

Contrarily, for $\lambda_4^{res} = 1556$ nm (see Figs. \ref{fig:distr}(g,h,i)), the maximum of the normalized E-field intensity becomes $|E|^2 = 2\times 10^4$ (see Fig. \ref{fig:distr}(i)) equivalent to an optical local field factor $f_{op}(x,y,z) = 141$. Unfortunately, the E-field confinement occurs outside the structure with a low intensity inside LN, which corresponds to the excitation of an almost air mode. The latter could be suitable to some specific applications like probing surface molecular modifications \cite{blin:aom13} but is not consistent with electro-optical-based applications. 
 
Knowing that for $\lambda_1^{res}$ the maximum of the E-field is confined solely in the membrane, we prefer to work with this mode that could be more robust towards the manufacturing imperfections . 

As in ref. \cite{calero:sr19}, the high slope value at half-width of the resonance could lead to high EO modulation strengths, which means a high sensitivity. Therefore, the latter is expressed as the minimum E-field required to spectrally notice the shift of the resonance. Generally, the sensitivity is defined as:
 \begin{equation}
 \textrm{Sensitivity} = \frac{|\Delta\lambda_{0}|}{| E_s|},
 \label{eq:sens}
 \end{equation}
 where $\Delta \lambda_{0}$ represent the spectral shift of the resonance dip for an applied electrostatic field $E_s$.
 \begin{figure}[h!]
	\includegraphics[width=130mm]{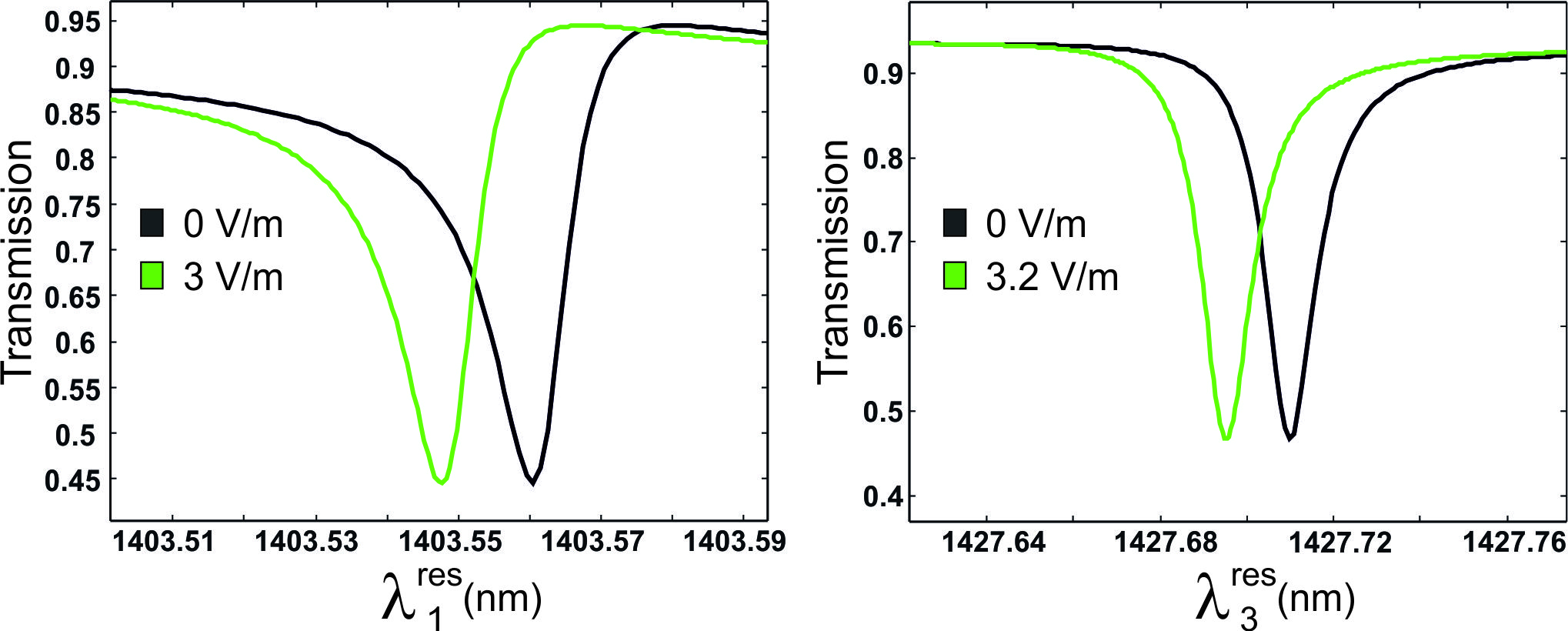}
	\centering
	\caption{Transmission spectra for the resonance $\lambda^{res}_1$ and $\lambda^{res}_3$ for the free structure (black lines) and when external E-fields are applied (green lines) with $E_s=3$ V/m and $E_s=3.2$ V/m.}
	\label{fig:sensi}
\end{figure}

To estimate this sensitivity, we perform numerical simulations in three steps: (a) The structure is first assumed to be free from any external E-field meaning that LN has a homogeneous refractive optical index and we calculate the transmission spectrum. (b) From the latter we perform a second FDTD numerical simulation in CW regime at $\lambda=\lambda_{res}$ to estimate, cell by cell, the optical field factor given by Eq. \ref{eq:fop}. (c) In the last step, after injecting these $f_{op}(x,y,z)$ values into Eq. \ref{eq:dnfop}, we carry out a new FDTD-simulation taking into account the modification of the refractive optical index of each FDTD cell and determine the transmission spectrum in the presence of an external E-field $E_s$ whose value is fixed.

\begin{figure}[h!]
	\includegraphics[width=110mm]{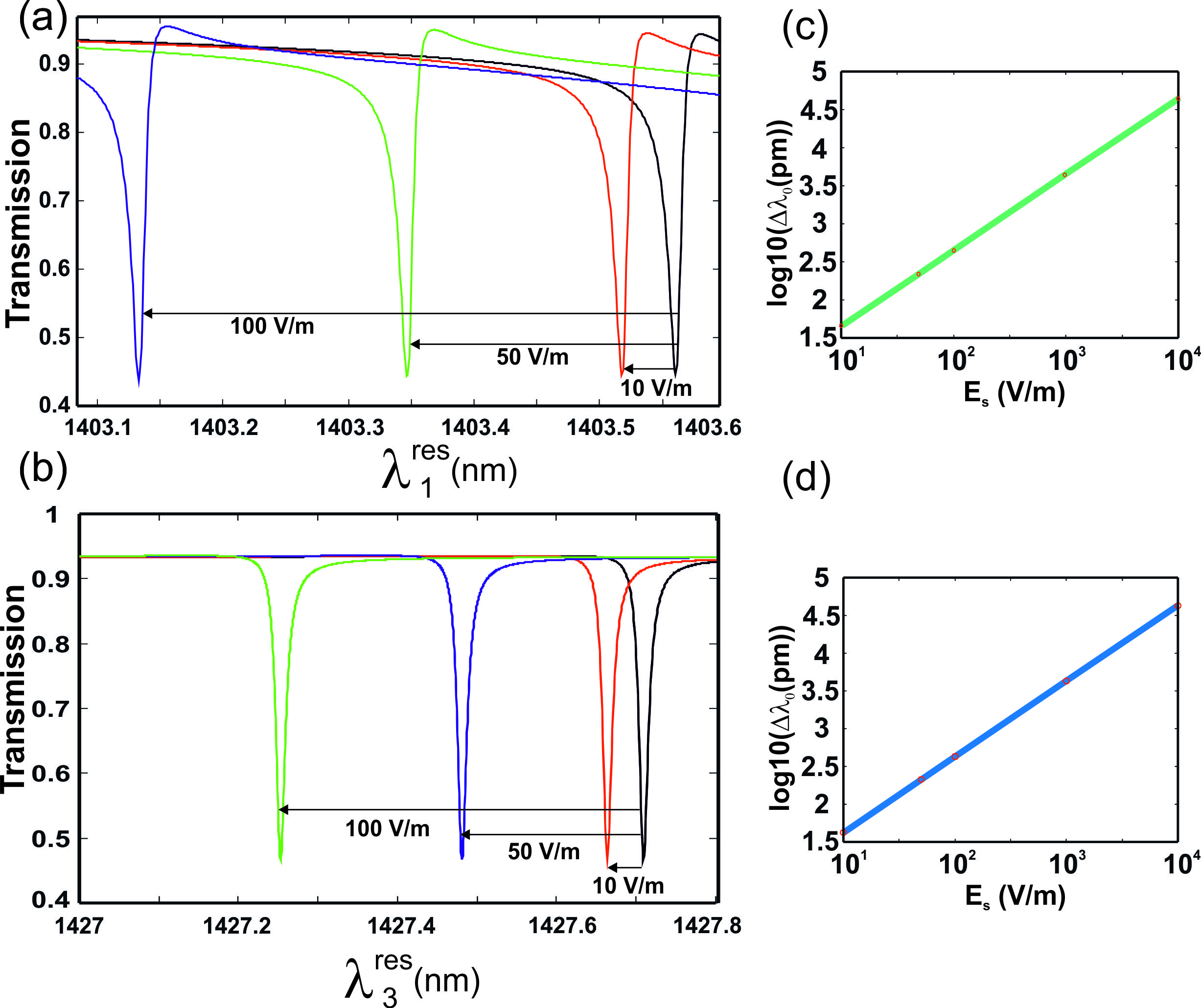}
	\centering
	\caption{The spectral position shift of (a)$\lambda_{res1}$ and (b)$\lambda_{res3}$ as function of the external E-field $E_s$. The variation of the spectral shift of (c)$\lambda_{res1}$ and (d)$\lambda_{res3}$ versus the applied electric field. It shows a linear behavior.}
	\label{fig:sensitivity}
\end{figure}

As a result, an E-field of around $3$ V/m is required to tune over a spectral range of width $11.9$ pm corresponding to the FWHM of the resonance at $\lambda_1^{res}$ and $E_s=3.2$ V/m for $\lambda_3^{res}$ over $14.0$ pm as shown in Fig. \ref{fig:sensi}. Experimentally, this shift is sufficient to be detected through the measurement of the transmission spectrum using a common low-price Optical Spectrum Analyzer (OSA) (such as MS9710B from Anritsu, which has a read resolution of $5$ pm\footnote{\url{http://www.smart-inc.com.tw/uploads/root/MS9710B_E1700.pdf}}
 ). Thus, if an Optical Spectrum Analyzer (OSA) with a high resolution of $0.1$ pm is employed (such as the BOSA 100 from Aragon Photonics which has an optical resolution of $0.08$ pm/$10$ MHz \footnote{\url{https://www.gophotonics.com/products/optical-spectrum-analyzers/aragon-photonics/58-667-bosa-100}}), we estimate to $25~\mu$V/m the minimum detected electric field.
\newline
 Fig. \ref{fig:sensitivity} shows the different spectra for different external E-field values ($0$, $10$ V/m, $50$ V/m, $100$ V/m) for the two resonances $\lambda^{res}_1$ in (a) and $\lambda^{res}_3$ in (b). For both cases, we plot on Figs \ref{fig:sensitivity}(c) and (d) the shift of the resonance as a function of the external E-field. As expected, the two behaviors are linear with quite similar slope values corresponding to the sensitivity of the configuration which is equal to $S=4$ pm.m/V. Let us note that this sensitivity is more than $160000$ times greater than the one obtained in ref. \cite{lipson:opex18} (see Fig. 3(a) of that paper) where a shift of $7$ pm/V is announced for electrodes separated by a gap of $3.5$ $\mu$m (electric field of almost $286$ kV/m).

\section{Conclusion}
In conclusion, we have designed and numerically studied by home-made FDTD codes the excitation of ultra-high Q-factor resonances up to $1.2 \times 10^6$ associated with the excitation of SPMs which are excited assuming an illumination with a small angle of incidence $\theta=0.5^\circ$. 
Furthermore, by exploiting such sharp resonances, we can develop an E-field sensor with a sensitivity of $4$ pm.m/V. It results from a local confinement of the electromagnetic field reaching a local optical field factor up to $245$. This kind of hybrid Fano-symmetry protected modes opens the way to a new generation of sensors assuming good quality manufacturing processes able to guarantee a high degree of symmetry of the fabricated structure. External E-fields as small as $25$ mV/m can then be detected with such device, making it possible to be used for medical investigations such as ECG.
\section*{Disclosures}
The authors declare no conflicts of interest.

Add references with BibTeX or manually.



\end{document}